\documentclass{aa} 
\usepackage{graphicx} 
\usepackage{supertabular}
\usepackage{natbib}
 
\begin{document} 
\newcommand{\inthms}[3]{$#1^{\rm h}#2^{\rm m}#3^{\rm s}$} 
\newcommand{\dechms}[4]{$#1^{\rm h}#2^{\rm m}#3\mbox{$^{\rm s}\mskip-7.6mu.\,$}#4$} 
\newcommand{\intdms}[3]{$#1^{\circ}#2'#3''$} 
\newcommand{\decdms}[4]{$#1^{\circ}#2'#3\mbox{$''\mskip-7.6mu.\,$}#4$} 
\newcommand{\tmb}{\mbox{T$_{\rm mb}$}} 
\newcommand{\OI}{\mbox{O\,{\sc i}}} 
\newcommand{\Htwo}{\mbox{H$_{2}$}} 
 
\hyphenation{Fe-bru-ary} 
 
\title{Theoretical HDO emission  
from low-mass protostellar envelopes} 

\author{B.Parise\inst{1}\thanks{Present address: Max-Planck Institut f\"ur Radioastronomie, Auf dem H\"ugel 69, 53121 Bonn, Germany}
 \and C.Ceccarelli\inst{2}  
 \and S.Maret\inst{3} 
} 
\institute{  
 CESR CNRS-UPS, BP 4346, 31028 - Toulouse cedex 04, France 
\and  
Laboratoire d'Astrophysique, Observatoire de Grenoble - 
 BP 53, F-38041 Grenoble cedex 09, France 
\and 
Department of Astronomy, University of Michigan, 500 Church Street, Ann Arbor MI 48109-1042, USA
 } 
 
\offprints{bparise@mpifr-bonn.mpg.de}   
 
\date{Received {\today} /Accepted {\today}} 
\titlerunning{Models of HDO emission} 
\authorrunning{Parise et al.} 
 
\abstract{ 
We present theoretical predictions of the rotational line  
emission of deuterated water in low-mass protostar collapsing envelopes. 
The model accounts for the density and temperature structure of the envelope,
according the inside-out collapse framework.
The deuterated water abundance profile is approximated by a step 
function, with a low value in the cold outer envelope and a higher value
in the inner envelope where the grain mantles evaporate.
The two abundances are the two main parameters of the modeling, along with the
temperature at which the mantles evaporate.
We report line flux predictions for a 30 and 5 L$_\odot$ source luminosity
respectively.
We show that ground based observations are capable to constrain 
the three parameters of the model in the case of bright low-mass protostars (L$>$10 L$_{\odot}$), and that no space based observations, like for example HSO observations, are required in this case. On the contrary, we show that the study of low-luminosity sources (L$<$10 L$_{\odot}$), assuming the same HDO abundance profile, requires too much integration time to be carried out either with available ground-based telescopes or with the HIFI instrument on board HSO. For these sources, only  the large interferometer ALMA will allow to constrain the HDO abundance.
} 
 
\maketitle 
 
\section{Introduction} 
 
Water is a key molecule in the study of star formation for four main reasons.
Since it has a large dipole, it has large spontaneous emission
coefficients and high critical densities, which makes it a 
powerful diagnostic tool.
Secondly, water is important in star forming regions for the role it plays 
in cooling the gas \citep{Neufeld93}, and, sometimes, also in heating it \citep[e.g. ][ hereinafter CTH96]{Ceccarelli96}.
Third, it is also an important molecule from the chemical point of view,
because in many circumstances it is the most abundant oxygen-bearing molecule,
and, hence, it influences the abundance of all the more complex ``trace''  
molecules; in other words, it influences the overall chemical composition 
of the gas \citep[e.g. ][]{Rodgers03}.
Finally, water is certainly the most abundant component of grain mantles too \citep[e.g. ][]{Gibb04}.
For all these reasons, it is extremely important to know its abundance
distribution in star forming regions.
Unfortunately, the interstellar rotational water lines are absorbed 
by the water vapour of the Earth atmosphere. Therefore, the observation 
of the water vapour in astrophysical sources requires out-of-atmosphere 
instruments, like those on the past satellites ISO, SWAS and ODIN, 
or the future satellite HSO, to be launched in 2007.

Fortunately enough, some rotational transitions of the deuterated form of 
water, HDO, are observable from ground based telescopes, and can be used,
under some circumstances, to probe the water content in astrophysical sources. 
Furthermore, HDO is an interesting molecule on its own, 
because it gives important
information on the water formation route on the one hand, and, on the other
hand, on the physical conditions of the observed region, either at present
or during the past.
If formed on the grain surfaces during a previous phase, for example,
and then released into the gas phase because of the mantle evaporation
due to the heating from a new born star, the measured HDO abundance would
bring the hallomark of the past history of water formation.
This is likely the case of what happens in solar type protostars.
After a pre-collapse cold phase where the grain mantles are likely formed,
either because of accretion of molecules onto the grain surfaces or because
of the synthesis of the molecules on the grain surfaces, the molecules are
released in the gas phase in the regions heated by the protostar radiation.
Two classes of molecules show evidence of this process:\\
i) Hydrogenated molecules like formaldehyde and methanol \citep{Ceccarelli00b,Schoier02,Schoier04,Maret04,Maret05} and complex organic 
molecules, observed in the so called ``hot cores'' \citep{Cazaux03,Bottinelli04a, Bottinelli04b,Kuan04}. 
These molecules are not observed in the cold gas with the large abundances
measured in the very inner regions of the envelopes surrounding solar
type protostars. 
Indeed, they are direct and indirect products of the grain mantle evaporation,
so that they are, in a way or another, linked to the pre-collapse phase.\\
ii) Deuterated molecules observed with extremely large abundances,
which can only be formed during the cold, dense and CO depleted
phase of the pre-collapse \citep{Roberts00a, Roberts03}. 
Notable examples are formaldehyde \citep{Ceccarelli98}, methanol
\citep{Parise02, Parise04}, hydrogenated sulfide \citep{Vastel03} and 
ammonia \citep{Lis02, vanderTak02}.\\
HDO belongs to the second class of molecules witnesses of the past
pre-collapse phase. And, it is, apparently, a very different beast from
that zoo.
Indeed, the above mentioned molecules show extremely large deuteration ratios
in low mass star forming regions:
singly deuterated molecules are more than 10\% of their H-counterparts,
whereas doubly and triply deuterated molecules can respectively be as high 
as 10\% and $\sim$ 1\%.
On the contrary, HDO is only a relatively small fraction of H$_2$O,
about 3\% at best \citep{Parise05}, and no D$_2$O has been
detected so far, at relatively low ratios to our knowledge (Cernicharo,
private communication).
The above relative small values agree indeed also with the non-detection
of solid HDO in protostars \citep[e.g. ][]{Dartois03, Parise03}.
Thus, there is evidence that the water deuteration follows a different
route with respect to the other molecules. In addition,
the study of the available observations in IRAS16293-2422 suggests that
the deuteration ratio is not the same in the inner and outer envelope:
3\% and $\le$0.2\% in the inner and outer envelope respectively
\citep{Parise05}.
Thus, in order to better understand this difference, one needs 
to be able to disentangle where the different emission comes from. 
The present study aims at giving a theoretical tool for the estimate of the
HDO abundance in protostellar envelopes, in the inner and outer regions,
and, possibly, the evaporation temperature,
and, consequently, at guiding the best observations to obtain those
informations.
Specifically, given that the satellite HSO will be launched in a relatively
near future (2007), a particular care will be devoted to discuss whether
observations with instruments on this satellite will be required for
deriving the HDO abundance and evaporation temperature
in low mass protostellar envelopes, against the observations
that can be obtained with ground based telescopes.

The article is organized as follows: we describe the adopted model and code
in \S 2, we present the HDO molecule as well as the instruments available
to observe its rotational spectrum in \S 3. We describe the general characteristic of the predicted spectrum 
in \S 4
and give the results in \S 5 for a relatively high (30 L$_{\odot}$) 
and low (5 L$_{\odot}$) luminosity case respectively. 
Finally, \S 6 contains a discussion
about the best lines to observe to derive the HDO abundance in the
inner and outer envelope, as well as the HDO ices evaporation temperature.
We also discuss the interest  of carrying space based versus ground based HDO 
observations to constrain these parameters.

\section{The model description} 

The model used to compute the HDO emission in collapsing envelopes
has been adapted from the original model developed by CHT96 to
predict the OI, CO and H$_2$O line emission, 
and successively modified to compute the H$_2$CO
line emission \citep{Ceccarelli03}. 
Here we describe very briefly the basic assumptions of the model.

The envelope density structure and dynamics are assumed to follow the  
``inside-out'' collapse picture \citep{Shu77}, 
for a spherical initial isothermal sphere undergoing collapse. 
In the outer envelope, not affected by the collapse yet,  
the molecular hydrogen number  
density distribution $n_{H_2}(r)$ is given by:  
\begin{equation} 
\label{static}n_{H_2}(r)=\frac{a^2}{2\pi \mu m_{H} G}r^{-2} \end{equation} 
$$=2.8\times 10^8  \left( \frac{a}{0.35 {\rm  ~km s}^{-1}} \right) ^2 r_{100AU}^{-2}  {\rm  ~cm}^{-3} ~$$
\noindent where $a$ is the sound speed, $m_{H}$ is the hydrogen mass, $\mu$ is the mean 
molecular mass in amu units, equal to 2.8,  
$r_{100AU}$ is the distance from the center in 100 AU units, and $G$ is 
the gravitational constant.  
In the inner collapsing regions the density is described by the 
free-fall solution:  
\begin{equation} 
\label{ffdens}n_{H_2}(r)=\frac 1{4\pi \mu m_{H}} 
\left(\frac{\dot M^2}{2GM_\star}\right)^{\frac 12} r^{-\frac 32}~ 
\end{equation} 
$$=1.2 \times 10^7 \left( \frac{\dot M_{-5}^2}{M_\star r_{100AU}^3} \right) 
^{\frac 12} {\rm ~cm}^{-3}~. ~~~~~~~~~$$

The free fall velocity is given: 
\begin{equation} 
\label{ffvel}v(r)=\left( \frac{2GM_\star}r\right) ^{\frac 12}~ 
=4.2 \left( \frac{M_{\star 1}}{r_{100AU}} \right) ^{\frac 12} 
{\rm km ~s}^{-1}~,  
\end{equation} 
where $\dot M$ is the mass accretion rate, related to the sound speed by  
\begin{equation} 
\label{accrrate}\dot M=0.975\frac{a^3}G  
=10^{-5}\left( \frac{a}{0.35 {\rm km~s}^{-1}} \right)^3 
{\rm ~M}_\odot~{\rm yr}^{-1}.  
\end{equation} 
$\dot M_{-5}$ is $\dot M$ in units of $10^{-5}$ M$_\odot$yr$^{-1}$,  
and $M_{\star 1}$ is the mass of the central object $M_\star$  
in units of 1 M$_\odot$. 
The gravitational energy of the collapsed mass is released radiatively as material falls 
onto the core radius $R_{\star}$, so that the luminosity of the central object is:
\begin{equation}
L_{\star} = \frac{GM_{\star}\dot M}{R_{\star}} = 22\left(\frac{M_{\star}}{M_{\odot}} \right)~ \dot M_{-5} ~R_{12}^{-1} ~L_{\odot}
\end{equation}
where $R_{12}$\,=\,R$_{\star}$/10$^{12}$ cm.  

The spherical symmetry is conserved through the collapse in this model. 
In this sense, the model gives accurate results only for that part 
of the envelope in which the spherical symmetry is a good approximation, 
i.e. probably for radii larger than a few tens of AUs \citep{Ceccarelli00a}. 
At smaller scales large deviations from the spherical symmetry 
are expected because of the presence of a circumstellar disk. 
 
The gas temperature is computed self-consistently by equating the cooling
and heating mechanisms at each point of the envelope. Details of these
calculations can be found in CHT96. What is important to notice here
is that the cooling depends on the abundance of the atomic oxygen, CO and 
H$_2$O, namely the main gas coolants across the envelope. 
Hence their abundances are hidden parameters of the model.
The resulting gas temperature tracks the dust temperature very closely 
across the envelope, except where the icy mantles evaporate. Because of 
the injection
of large quantities of water in the gas phase and the consequent enhancement
of the gas cooling, the gas and dust (thermally) decouple in a 
relatively small region after the icy mantles evaporation 
\citep[see also for example][]{Maret02}.
 
\citet{Ceccarelli00a} and \citet{Maret02} demonstrated that the gas temperature differs 
by no more than 10\% from the dust temperature, and this occurs in a relatively
small region just next to where the ices sublimate.
There is some uncertainty in the water and oxygen abundances in protostars,
for derivations of the H$_2$O abundance has been obtained in two sources only,
IRAS 16293 and NGC1333-IRAS4 \citep{Ceccarelli00a, Maret02}.
However, these uncertainties do not affect much the gas temperature derivation,
for in the outer envelope the gas is cooled mostly by CO\footnote{The CO lines are optically thick and, therefore, the cooling function does not depend on the CO abundance unless it decreases by a factor of 100 with respect to the standard value.}, and the inner is cooled
by  the sublimated water. At the same time, in the inner region the heating of the gas is dominated by the collisional de-excitation of the water molecules, photopumped by the photons emitted by the innermost warm dust (CHT96), so that, in the end, the coupling between the dust and gas is always ensured. As a result, the HDO computed line intensity are rather
insensitive to these uncertainties.

The original CHT96 model is time-dependent, but the authors demonstrated that
the physical structure and water chemistry can be computed at each time independently
on the previous history, once the luminosity and the mass of the protostar are known.
The reason for that is the following. In the outer cold envelope the abundance of the water is
practically constant with time, whereas in the inner warm envelope the water abundance
is set by the ice evaporation, except in the very inner regions where the temperature
exceeds 250\,K. Since the sizes of the evaporation region are set by the temperature and
the density profile, for practical applications to real sources one needs to input them only. In
addition to that, the velocity field is also needed to compute the line intensity, and this can be derived as well by observations.

For the standard case described in section 4, we considered the case of IRAS 16293, 
whose velocity, density and temperature profiles have been derived by several 
authors from actual observations. The computations shown in Fig \ref{linetest}-\ref{temp} adopted the profile derived in \citet{Ceccarelli00a}.
We also studied the case of a low luminosity source, and, at this scope, we used the
velocity, density and temperature profiles derived for the source L1448mm
by \citet{Maret04}.

Finally, we approximate the HDO abundance profile with a step function. 
In the outer envelope the HDO abundance, x$_{\rm cold}$, is 
relatively low and similar to that observed in molecular clouds. 
In the innermost regions of the envelope, where the dust reaches the  
evaporation temperature T$_{\rm evap}$,
the HDO abundance, $x_{\rm warm}$, jumps to larger values. 
These last three parameters --- $x_{\rm cold}$, $x_{\rm warm}$ and 
T$_{\rm evap}$ --- are the three free parameters of the present modeling,
and they will be varied to study their influence on the HDO line emission.

Low-mass protostars exhibit energetic molecular outflows, where the gas can be very hot 
(up to 2000\,K, as testified by the H$_2$ rovibrational emission observed in several outflows). This gas
could in principle contribute to the HDO line emission, so that it is important to evaluate whether
the outflow emission can be comparable or even larger than the HDO emission from the envelope. 
From a theoretical point of view, since molecular deuteration decreases exponentially with increasing gas temperature, the HDO abundance is expected to be very low in the warm gas of molecular outflows \citep[see also][]{Bergin99}. To support these theoretical expectations, observations of outflows on several 
low-mass protostars have revealed no HDO emission, at a level at least 10 times 
fainter than towards the protostar (IRAS16293, \citealt{Parise05}; NGC1333-IRAS2 and IRAS4A, \citealt{Caux05}).
We therefore decided to consider in 
the present study only the envelope contribution to the HDO emission.

All the line fluxes are computed for a source at a distance of 160 pc. 
The model computes the level population of the first 
34 levels of the HDO. 
The molecular data are from the JPL catalogue 
({\it http://spec.jpl.nasa.gov/home.html}), and 
the collisional coefficients are from \citet{Green89},  
for He-HDO collisions between 50 and 500 K, scaled for collision with H$_2$. 
Note that all abundances are reported here with respect to H$_2$. 
 
\section{The HDO rotational line spectrum}

\subsection{The energy levels}
HDO is a planar assymetric top molecule. Its dipole lies along its $a$ and $b$ axies, with 
dipole moments $\mu_a$\,=\,0.657 and $\mu_b$\,=\,1.732 Debye \citep{Clough73}. The allowed transitions are thus $a$-type and $b$-type transitions~:
$$ \Delta{\rm K}_a = 0 ~~{\rm and}~~  \Delta {\rm K}_c = \pm 1 $$
$$ {\rm or} $$ 
$$ \Delta{\rm K}_a = \pm 1 ~~{\rm and}~~  \Delta {\rm K}_c = \pm 1~. $$

\begin{figure}[!h]
\includegraphics[width=0.54\textwidth]{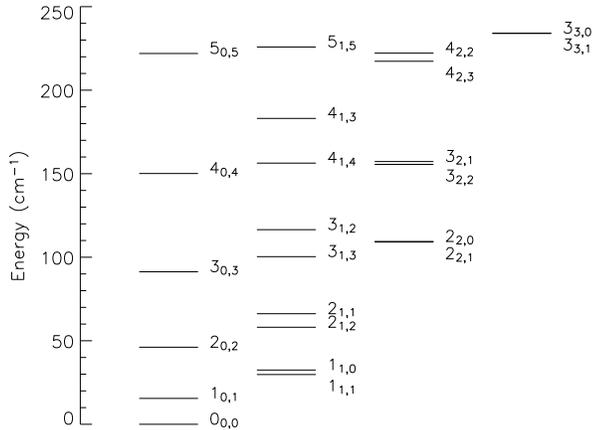}
\caption{First energy levels for the HDO molecule. }
\label{level}
\end{figure}

The first energy levels for HDO are presented in Fig.~\ref{level}.

The list of the transitions between the first 34 levels is reported in Table \ref{fluxtest}.

\subsection{The observing instruments}

We detail in this section the instruments that operate (or will operate in the near-future) in the 
frequency bands interesting for the HDO observations. 
\smallskip

{\noindent a) Ground telescopes}
\smallskip

Several ground based facilities allow to observe the HDO rotational spectrum transitions.\\
The IRAM (Institute for Millimetre Radioastronomy) 30\,m telescope (Pico Veleta, Spain) operates between 80 and 281\,GHz.\\
The 15\,m JCMT dish ( James Clerk Maxwell Telescope, Mauna Kea, Hawaii) observes in the 230, 345, 450, and 650 GHz frequency bands. The 800\,GHz instrument is no longer available, but this frequency range will be observable with APEX.\\
The 10\,m CSO dish (Caltech Submillimeter Observatory, Mauna Kea, Hawaii) operates roughly in the same frequency bands as JCMT.\\
The 15\,m APEX dish (Atacama Pathfinder EXperiment, Chajnantor, Chile) is being built on the ALMA site. It will allow observations in the 230, 345, 650 and 850 GHz bands.
\smallskip

{\noindent b) Airborne instruments: SOFIA}
\smallskip

The 2.5\,m telescope of the Stratospheric Observatory for Infrared Astronomy, on board a Boeing 747, will be equipped with several interesting instruments for HDO observation. In particular, CASIMIR (CAltech Submillimeter Interstellar Medium Investigations Receiver), will observe in the 500-2100 GHz band and GREAT (German REceiver for Astronomy at Terahertz frequencies) in the 1.5-5 THz. 
\smallskip

{\noindent c) The HSO satellite}
\smallskip

The HIFI (Heterodyne Instrument for the Far-Infrared) instrument on board Herschel Space Observatory (launch planned in 2007) will allow observations in the 480-1250 and 1410-1910 GHz bands.

\section{Model results} 
 
We first report the computed HDO line spectrum of  
a test case, to illustrate the general characteristics 
of the predicted spectrum. 
Then, in the following paragraphs we discuss thoroughly  
how the HDO line spectrum varies with the three free parameters of the model, 
namely the  HDO abundance in the outer cold envelope $x_{\rm cold}$, and  
in the inner warm envelope $x_{\rm warm}$, and the evaporation
temperature T$_{\rm evap}$. 

\begin{table}[!h] 
\centering
\caption[]{Values of the parameters adopted for the test case. The upper
part reports the values of the three free parameters of the model, whereas
the lower part reports the value of the hidden parameters.} 
\begin{tabular}{cc} 
\hline 
\hline
\noalign{\smallskip}
Parameter & Value \\ 
\noalign{\smallskip}
\hline 
\noalign{\smallskip}
$x_{\rm warm}$ & $1 \times 10^{-7}$ \\ 
$x_{\rm cold}$ & $1.5 \times 10^{-10}$ \\ 
T$_{\rm evap}$ & 100~K\\
\noalign{\smallskip}
\hline
\noalign{\smallskip}
M$_\star$ & 0.8 M$_\odot$ \\ 
\.M  & $3 \times 10^{-5}$ M$_\odot$yr$^{-1}$\\ 
x(CO) & $1 \times 10^{-4}$ \\ 
x(O)  & $2.5 \times 10^{-4}$ \\ 
x$_{\rm cold}$(H$_2$O) & $5 \times 10^{-7}$ \\ 
x$_{\rm warm}$(H$_2$O) & $3 \times 10^{-6}$ \\ 
\noalign{\smallskip}
\hline 
\end{tabular} 
\label{testcase} 
\end{table} 

The parameters adopted in the test case are reported in Table \ref{testcase}.
The values of the hidden parameters  are the same of the test case 
shown in \citet{Ceccarelli03} for discussing the H$_2$CO line emission, and
have been derived from several observations of IRAS16293-2422. 
The $x_{\rm cold}$, $x_{\rm warm}$ and T$_{\rm evap}$ parameters
have also been derived by the analysis of the HDO observations
towards IRAS16293-2422 \citep{Parise05}.
In the following all fluxes are given in erg\,s$^{-1}$\,cm$^{-2}$ 
and are the integrated emission over the entire envelope for a source at 160 parsecs. 
Although the conversion in K\,km\,s$^{-1}$ would have been more convenient   
for the observers, the signal in K\,km\,s$^{-1}$ depends  
on the beam of the telescope used for 
the observation and how it matches the predicted extent of the emission. 
We will give the K\,km\,s$^{-1}$ signal for a few lines, when considering
the application to observations with ground versus space based instruments
(\S 5).

\normalsize

\begin{figure}[tb] 
\includegraphics[angle=0,width=1.\columnwidth]{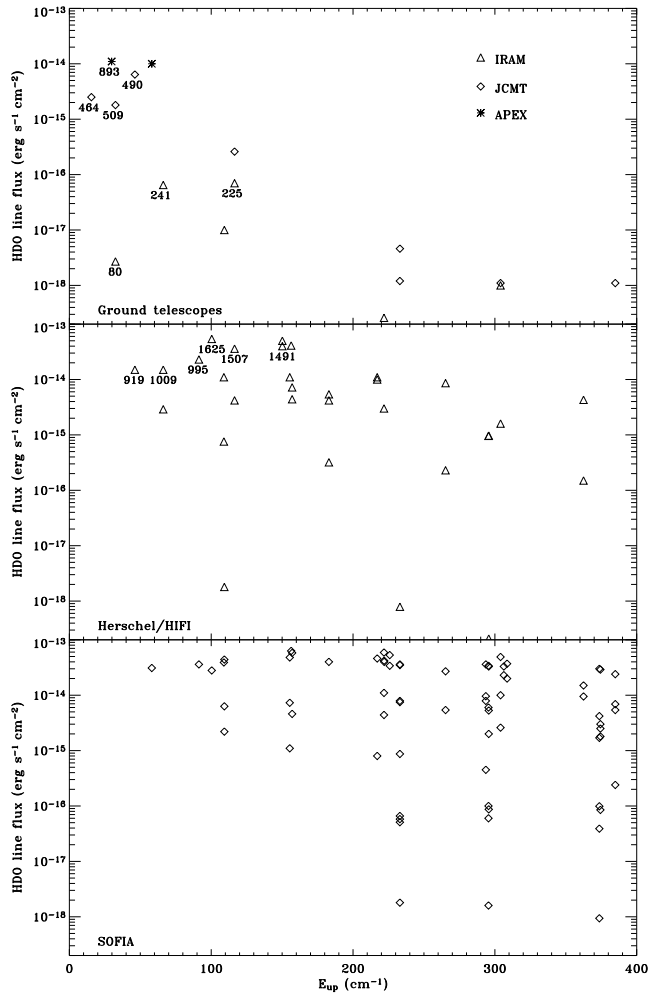} 
\caption{HDO line fluxes predicted for the test case IRAS16293$-$2422 versus the upper level energy of the transition. The upper panel presents the transitions observable from the ground, the middle panel the transitions observable with HSO/HIFI and the lower panel the transitions observable with SOFIA. The frequencies of the transitions that are going to be studied in more detail are indicated in GHz. } 
\label{linetest} 
\end{figure}

Table \ref{fluxtest} reports the predicted fluxes of the HDO lines for 
the test case, and Fig. \ref{linetest} shows the line fluxes as a function of the upper 
energy of the transition. 

The transitions have been labeled according to the available observing facility. It should be noted 
that the lowest lying levels are observable from the ground and that space borne observations become competitive for upper energies greater than 50\,cm$^{-1}$.

\begin{figure}[tb] 
\includegraphics[angle=0,width=1.\columnwidth]{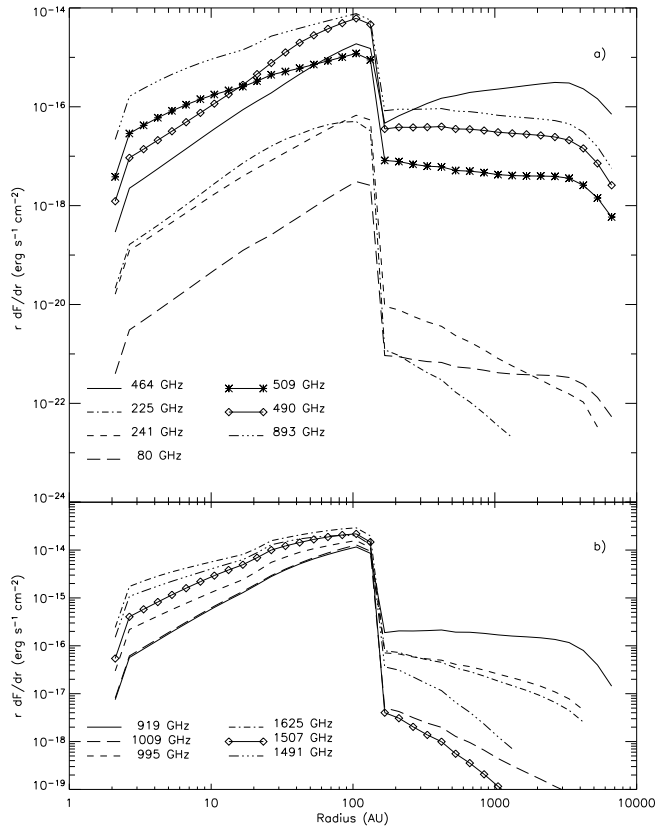} 
\caption{Contribution to the line intensity as function of 
the distance from the central source, for seven HDO lines
observable from the ground (upper panel) and six from space based
telescopes (lower panel).} 
\label{profiles} 
\end{figure} 

Fig. \ref{profiles} shows the contribution to the emission from  
different shells of the envelope for several transitions with different 
upper energy levels, observable with ground-based telescopes (upper panel)
or with Herschel-HIFI (lower panel).
Note that the 893\,GHz line has been included in the upper panel as 
it will be observable with the APEX facility.
Some of the lines have been observed towards IRAS16293$-$2422 using
the JCMT and the IRAM 30\,m telescopes \citep[80, 225, 241 and 464\,GHz,][]{Parise05}.
The other ones have been chosen because they are potentially bright 
lines (Fig. \ref{linetest}), spanning upper energy levels up to 150\,cm$^{-1}$
($\sim$\,200\,K). This figure illustrates that the relative contribution 
to the line intensity from the inner region (r $\leq 150$ AU) increases 
with increasing upper level energy of the transition. 
While all lines intensities are dominated by the emission in the
inner region (where the abundance is about 3 orders of magnitude higher),
the lines with low upper level energy exhibit a relatively
high contribution from the cold outer envelope.
 
\subsection{Varying the parameters} 
 
In this paragraph, we show how the HDO line spectrum changes varying 
the three free parameters of the model,  
namely the HDO abundance in the outer and inner parts of the envelope, 
$x_{\rm cold}$ and $x_{\rm warm}$  respectively, and the mantle 
evaporation temperature T$_{\rm evap}$. 
 
\medskip 
\noindent 
{\it a) $x_{\rm cold}$ and $x_{\rm warm}$} 
 
\begin{figure} 
\includegraphics[angle=0,width=1.\columnwidth]{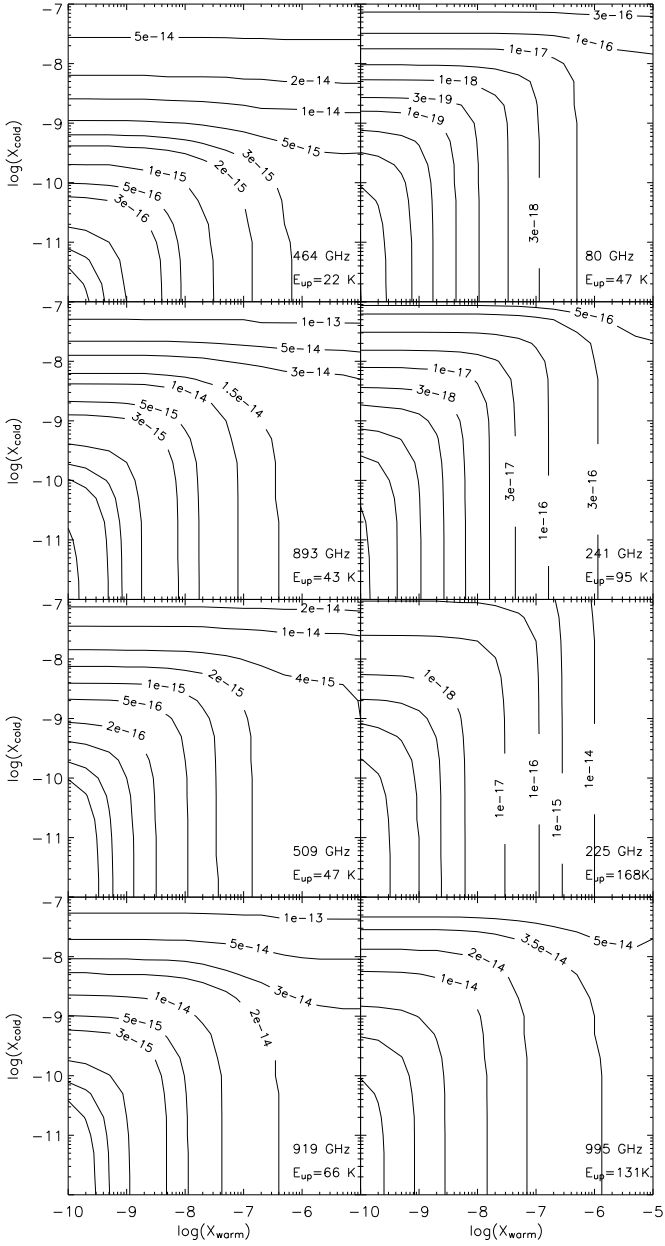} 
\caption{Line intensity  
as a function of the HDO abundance in the outer and inner regions 
of the envelope, $x_{\rm cold}$ and $x_{\rm warm}$ respectively, 
for the eight HDO transitions at 464, 893, 509, 80, 241, 225, 919 and 995  
GHz respectively. The first six lines are observable from the ground, whereas the last 
two ones will be observable with the Herschel telescope. 
Line fluxes are in erg\,s$^{-1}$\,cm$^{-2}$. 
In these calculations the mantle evaporation temperature is 100\,K. 
} 
\label{coolvswarm} 
\end{figure} 

Fig. \ref{coolvswarm} shows the line intensities of eight HDO lines 
(including two lines observable with Herschel-HIFI)
for $x_{\rm cold}$ varying from $10^{-7}$ to $10^{-12}$ 
and $x_{\rm warm}$ varying from $10^{-5}$ to $10^{-10}$. 
As expected, the intensities of the lower level transitions are mostly sensitive to the 
emission from the cold part of the envelope, as long as the outer abundance 
of HDO is not too low compared to the inner abundance. For instance
the 464\,GHz line is sensitive to $x_{\rm cold}$ even for inner abundances up to 
two orders of magnitude larger. On the contrary, the higher 
the upper energy, the more the transition becomes dependent on the 
inner parts of the envelope.

The figures also show that the low lying lines are much less sensitive
in general to the variation of $x_{\rm warm}$ (a factor 50 at most varying $x_{\rm warm}$
by four orders of magnitude) with respect to the high lying lines.
The latter vary almost linearly with $x_{\rm warm}$, so they are clearly better
suited to constrain this parameter than the low lying lines.
On the other hand, $x_{\rm cold}$ is better constrained by the low lying lines,
and particularly by the 464 GHz one, which shows the largest variation,
nearly 2 orders of magnitude for a change of $x_{\rm cold}$ of two orders of
magnitude.

 \medskip 
\noindent 
{\it b) T$_{\rm evap}$ } 

\begin{figure} 
\includegraphics[angle=0,width=1.\columnwidth]{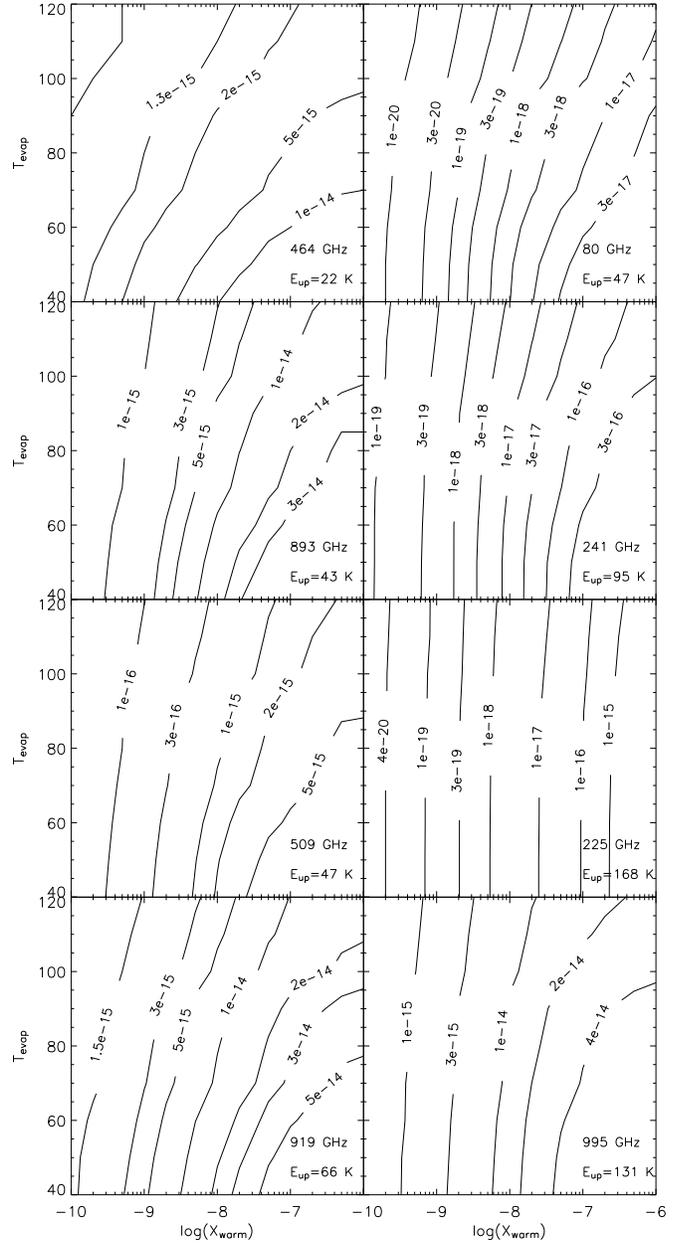} 
\caption{Predicted flux for the eight transitions depending on the assumed evaporation and inner HDO abundance for an outer abundance $x_{\rm cold}$ fixed to 2$\times$10$^{-10}$. } 
\label{invstemp} 
\end{figure} 

\begin{figure} 
\includegraphics[angle=0,width=1.\columnwidth]{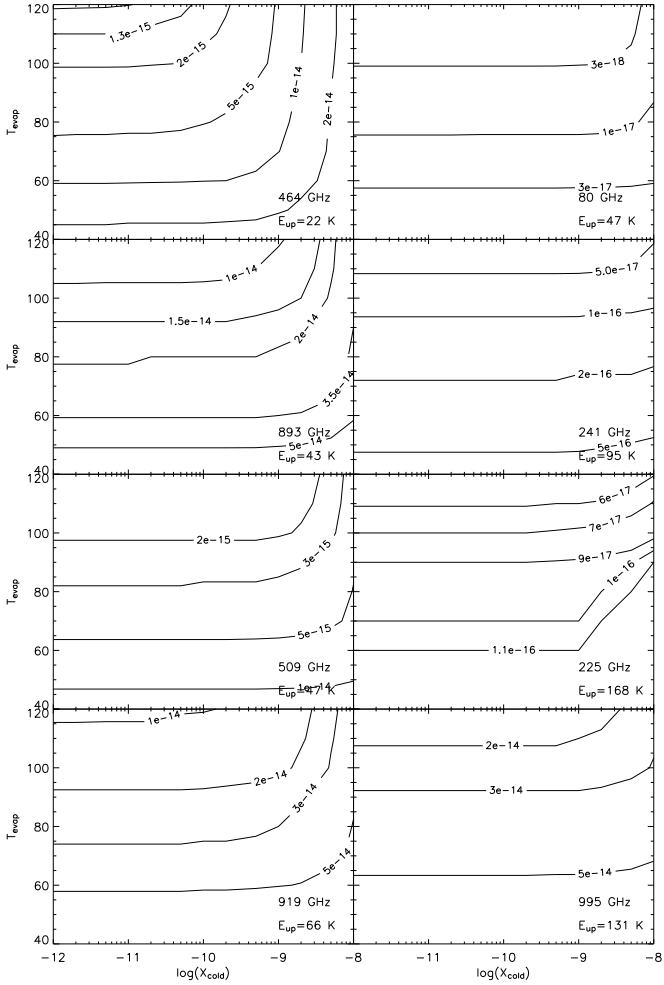} 
\caption{Predicted flux for the eight transitions depending on the assumed evaporation and outer HDO abundance for an inner abundance $x_{\rm warm}$ fixed to 10$^{-7}$. } 
\label{outvstemp} 
\end{figure} 

Fig. \ref{invstemp} and \ref{outvstemp} present the predicted fluxes 
depending on the evaporation temperature and respectively the inner 
and the outer abundances. The higher the evaporation temperature, 
the higher the inner abundance must be to reproduce the same flux. Indeed,
the higher the evaporation temperature, the less extended the inner region. 

The 225\,GHz line is mostly independant of the evaporation temperature in the 
studied range (Fig \ref{invstemp}). Indeed this line has the highest upper energy ($\sim$~170\,K) and is likely to be mostly excited in the innermost region of the envelope. Thus it  does not depend on the evaporation temperature, as long as this temperature is much lower than 170\,K, nor on $x_{\rm cold}$. This line is thus only sensitive to the inner abundance.

\begin{figure} 
\includegraphics[angle=0,width=1.\columnwidth]{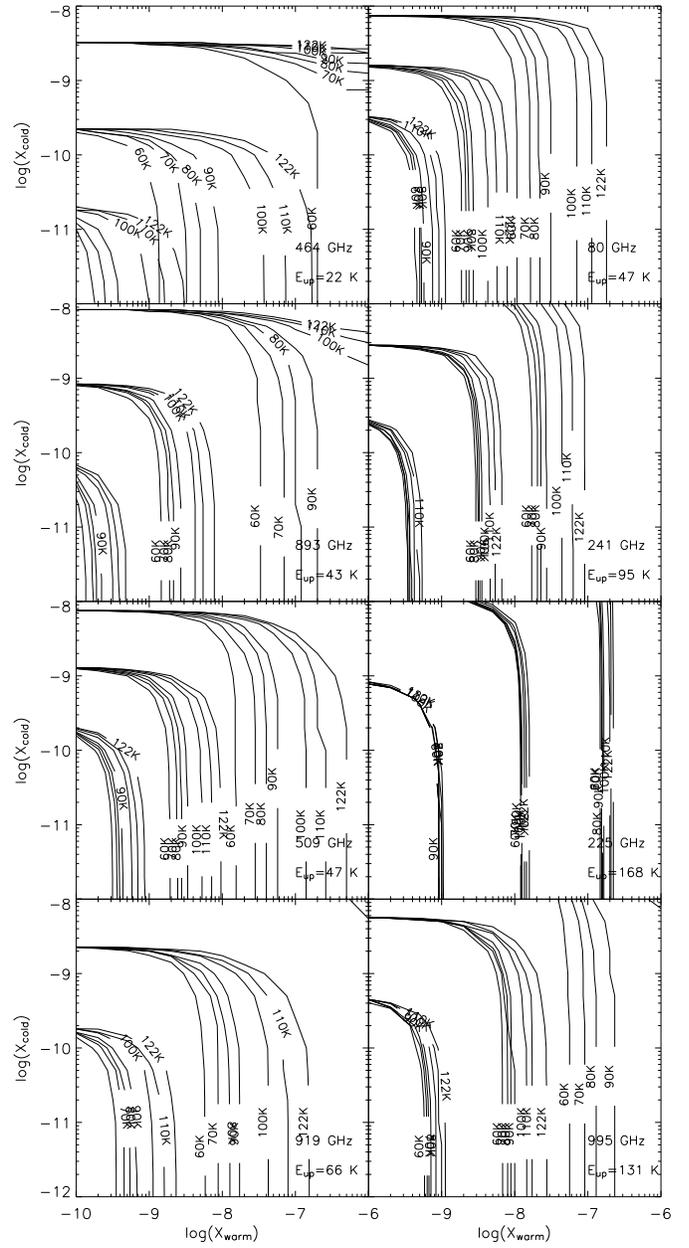} 
\caption{Evaporation temperature dependence in a $x_{\rm warm}$-$x_{\rm cold}$ diagram.
For each transition, the three sheaves correspond to the contour for three different values of the 
line flux. The different curves for each sheaf correspond to different values of the evaporation temperature. A narrow sheaf implies that the uncertainty on the evaporation temperature value will not affect much the estimation of $x_{\rm warm}$ and $x_{\rm cold}$. }
\label{temp} 
\end{figure} 

Another way to understand the dependance of the emission versus the evaporation temperature is to plot the regions spanned when varying the evaporation temperature for a fixed emission flux in a $x_{\rm cold}$-$x_{\rm warm}$ diagram (Fig. \ref{temp}).

As expected, we can notice that for low inner HDO abundances, the effect of the evaporation temperature is negligible, but curves begin to separate for high $x_{\rm warm}$. The uncertainty on the evaporation temperature mostly translates into an uncertainty in $x_{\rm warm}$. A remarquable point here is again that the 225\,GHz emission is independent of the temperature.

Fig. \ref{invstemp} shows that the line fluxes depend much more on x$_{\rm warm}$ than on
T$_{\rm evap}$. Actually, based on Fig. \ref{invstemp} and \ref{outvstemp}, it seems very difficult to find a way to constrain T$_{\rm evap}$.
However, Fig. 7 sheds some lights, and gives a sense of how the two parameters
can be likely constrained.
The 225 GHz line flux in practice can be used to estimate $x_{\rm warm}$ (Fig. \ref{outvstemp} and
\ref{temp}), irrespectively of T$_{\rm evap}$, whereas one can use, for example, the 464, the 80 and/or
the 509 GHz line fluxes to estimate T$_{\rm evap}$, knowing $x_{\rm warm}$, from Fig. \ref{temp}.
Evidently, it is not possible to easily disentangle the three parameters,
but the above analysis suggests that a full modeling can possibly succeed.

\section{Applications}\label{applications} 
 
\subsection{IRAS16293-2422}
In this section we try to give some practical recipes to derive, 
in first approximation, the values of the parameters of the model 
from actual observations. 
However, we emphasize that the described method will just give  
some approximative estimate of the parameter values, and that a full 
modeling of the source is necessary to derive more precise 
estimates. 
 
As a practical example, we will use the data obtained towards IRAS16293-2422 
\citep{Stark04, Parise05}
and we will derive the approximate values of $x_{\rm cold}$  
and $x_{\rm warm}$. We will then compare these results with those obtained 
with full modeling the HDO emission from this source, as described in 
\citet{Parise05}. 
 
First, to facilitate the comparison, 
Table \ref{conversion} reports the conversion factor of the line 
intensity in erg s$^{-1}$ cm$^{-2}$ into the velocity integrated  
main beam temperature T$_{\rm mb} \Delta$v observed at the  
IRAM and JCMT telescopes respectively, for the five lines observed towards IRAS16293$-$2422 \citep{Parise05}.
{\it Note that this is simply a conversion table, which does not take 
into account 
the convolution of the predicted emission with the telescope beam.} 
This is strictly valid for a point-like source with respect to  
the telescope beam sizes. 
The reader should be aware that the convolution with the beam  
can introduce large factors of difference, if the emission originates in a 
region larger than that encompassed by the used beam, as it may be the case 
for low lying lines. 
In those cases, in order to use the plots of this article, the observer 
should take care to integrate the emission over the entire emitting region. 

\citet{Parise05} performed a full analysis of the HDO emission towards IRAS16293$-$2422.
We describe however how the tools we have given here allow to approximately derive the 
envelope parameters. According to Figure \ref{temp}, the emission of the 225\,GHz line, as well as the upper limit on the $x_{\rm cold}$ derived from the 464\,GHz line, do not depend 
on the evaporation temperature. We can then use the Figure \ref{coolvswarm} (computed for an evaporation temperature of 100\,K) to derive an upper limit on $x_{\rm cold}$ from the 464\,GHz line. \citet{Parise05} measured a line intensity of 5.5\,K\,km\,s$^{-1}$. Using the conversion factor given in Table \ref{conversion}, this corresponds to 2.2$\times$10$^{-15}$ erg\,s$^{-1}$\,cm$^{-2}$. We deduce from Figure \ref{coolvswarm} that $x_{\rm cold}$ is lower than a few 10$^{-10}$ to 10$^{-9}$. The 225\,GHz line emission then allows to constrain the value of $x_{\rm warm}$ (independantly of the evaporation temperature). Indeed, converting the intensity observed by \citet{Parise05}, we find 7$\times$10$^{-17}$ erg\,s$^{-1}$\,cm$^{-2}$, which with the condition x$_{\rm cold}$ lower than 10$^{-9}$, constrains the abundance of HDO in the warm envelope to a value around 10$^{-7}$. The full modelling indeed leads to $x_{\rm warm}$=(1$\pm$0.3)$\times$10$^{-7}$, and to a 3$\sigma$ upper limit on $x_{\rm cold}$ of 10$^{-9}$ \citep{Parise05}.

\begin{table*}[tb] 
\begin{center}
\begin{tabular}{ccccccc} 
\hline 
\hline
\noalign{\smallskip}
Transition & Frequency & Telescope & Conversion coeff
 & \multicolumn{3}{c}{IRAS16293-2422}\\ 
           &  (GHz)    &            & 1\,K\,km\,s$^{-1}$ (T$_{\rm mb} \Delta$v)   & (K km s$^{-1}$) & (K km s$^{-1}$) & (K km s$^{-1}$)\\ 
 & & & (erg s$^{-1}$ cm$^{-2}$) & Conversion$^a$ & Convolution$^b$ & Obs$^c$ \\
\noalign{\smallskip}
\hline 
\noalign{\smallskip}
 $1_{1,0}-1_{1,1}$ & 80.578  & IRAM   & 1.4$\times$10$^{-17}$ & 0.19  & 0.21 & 0.40  \\ 
 $3_{1,2}-2_{2,1}$ & 225.897 & IRAM   & 4.1$\times$10$^{-17}$ & 1.7   & 1.8  & 1.7   \\ 
 $2_{1,1}-2_{1,2}$ & 241.561 & IRAM   & 4.4$\times$10$^{-17}$ & 1.5   & 1.5  & 2.0   \\ 
 $2_{2,0}-3_{1,3}$ & 266.161 & IRAM   & 4.9$\times$10$^{-17}$ & 0.20  & 0.22 & 0.21  \\ 
 $1_{0,1}-0_{0,0}$ & 464.924 &  JCMT & 4.0$\times$10$^{-16}$ & 6.3   & 5.4  & 5.5   \\ 
 \noalign{\smallskip}
\hline 
\end{tabular} 
\caption[]{Conversion factors of the line 
intensity in erg s$^{-1}$ cm$^{-2}$ into the velocity integrated 
main beam temperature 
T$_{\rm mb} \Delta$v observed at the IRAM and JCMT telescopes respectively, 
for the five lines observed towards IRAS16293$-$2422 \citep{Parise05}.
Last column quotes the T$_{\rm mb} \Delta$v 
observed towards IRAS16293-2422 \citep{Parise05}.$^a$\,Predicted flux 
from the model simply converted into K km s$^{-1}$ using the conversion factor.
$^b$\,Predicted flux derived using an accurate convolution of the signal with the
telescope beam.} 
\label{conversion} 
\end{center}
\end{table*} 
 
Table \ref{conversion} compares the predicted fluxes, derived from the model by 
both using the simple conversion flux for a point-like source and an accurate
convolution with the telescope (IRAM or JCMT) beam, to the observations 
performed towards IRAS16293$-$2422 \citep{Parise05}.
As expected the simple conversion works reasonably well for the high lying lines,
as they originate from the small inner envelope (small enough compared to the beam
size to be approximated by a point). The accuracy of the conversion for the ground
line at 464\,GHz is less good as some of the emission originates from the outer 
envelope, which is affected by the convolution of the telescope beam. Nevertheless
the simple conversion is still accurate to 15\% as for this testcase, the HDO
abundance is three orders of magnitude lower in the outer envelope compared to the
inner envelope. The conversion would become critical when the abundance in the cold 
envelope is not low compared to the warm contribution.  
 
As a final remark, the plots of Fig. \ref{coolvswarm}, \ref{invstemp} and \ref{outvstemp} can be used 
with single dish observations without the need of convolution, 
for the signal, originating in the warm part of the envelope,  
is likely to be encompassed by the telescope beam. 
 
\subsection{The low luminosity source case: L1448mm}
 
The previous predictions are of course dependent on the physical structure considered 
for the source. As an example, we also plotted the contours for a low-luminosity source, 
L1448mm, for which physical characteristics can be found in \citet{Jorgensen02} or \citet{Maret04}. The results are presented in Appendix \ref{l14}. The contours roughly look the same, 
except that the emission is weaker in L1448mm.

\subsection{Are space based telescope observations absolutely necessary towards low-mass protostars ?}

We saw in the previous sections that the observation of the studied lines from the ground allows to perfectly constrain the inner HDO abundance. However, the outer HDO abundance can only be constrained using lower energy transitions. Among the studied lines, only the 464\,GHz can provide, in  some specific conditions (namely when the outer abundance is not too low compared with the inner abundance), some information on $x_{\rm cold}$. Unfortunately, the transitions that will be observable from space are at higher energy (see Fig. \ref{linetest}). They will hence only provide 
information on the inner abundance, which is already correctly constrained with the transitions observed with ground based telescope. Moreover observations from the ground, when possible, have to be preferred owing to the better spatial resolution. 
Table \ref{int_time} lists the integration time required to observe some selected transitions with a peak signal to noise of 5, at a resolution of 0.5\,km/s, with the various instruments IRAM 30\,m, JCMT and HSO/HIFI. Note that in each case, only one receiver was assumed (single polarization mode) so that the integration time could in principle be lowered by a factor of $\sqrt 2$ if the observations are carried out in double polarization mode. An average 
of 2\,mm of water vapor has been assumed at the IRAM 30\,m, where the average elevation of the source is 
25$^{\circ}$ and 70$^{\circ}$ respectively for IRAS16293 and L1448mm. Weather band 2 was assumed at JCMT, with a zenith angle of 40 and 20 respectively. An efficiency factor of 0.125 was assumed for HIFI, as predicted in chopping mode.

\begin{table}
\begin{tabular}{llll}
\noalign{\smallskip}
\hline
\noalign{\smallskip}
Transition & IRAM 30m & JCMT & HSO/HIFI \\
\noalign{\smallskip}
\hline
\hline
\noalign{\smallskip}
\multicolumn{4}{l}{IRAS 16293 (30 L$_{\odot}$)}\\
\noalign{\smallskip}
\hline
\noalign{\smallskip}
80 GHz & 5.6 hours & & \\
225 GHz & 15 min & & \\
241 GHz & 30 min & & \\
464 GHz & & 29 min & \\
490 GHz & & 21 min & 7 min \\
509 GHz &  & 34 hours$^*$ & 104 min \\
893 GHz & & & 16 min \\
995 GHz & & & 5 min \\
\noalign{\smallskip}
\hline
\noalign{\smallskip}
\multicolumn{4}{l}{L1448mm (5 L$_{\odot}$)}  \\
\noalign{\smallskip}
\hline
\noalign{\smallskip}
80 GHz & 3500 hours & & \\
225 GHz & 110 hours & & \\
241 GHz & 227 hours & & \\
464 GHz & & 185 hours  & \\
490 GHz & &132 hours  & 73 hours\\
509 GHz & & 1562 hours$^*$  & 471 hours \\
893 GHz & & & 59 hours \\
995 GHz & & & 16 hours \\
\noalign{\smallskip}
\hline
\noalign{\smallskip}
\end{tabular}\\
$^*$ The sensibility drops on the edge of the band.
\caption{Integration time required to detect the HDO lines with a peak signal over noise of 5 (see text). Note that only one receiver was assumed, so in principle those times can be lowered by a factor $\sqrt{2}$ by using double polarization setup.}
\label{int_time}
\end{table}

These predictions show that for sources as bright as IRAS16293, observations can be carried out 
from the ground, and no Herschel time is thus necessary. But in the case of a low-luminosity 
protostar, where the same abundance profile of HDO was assumed, integration times start to become prohibitive from the ground. The 225 and 241 GHz lines may remain observable, using double polarisation receivers. Note that the integration time required with HIFI is prohibitive, the  
only reasonable transition being the 995 GHz. Thus, in the case of low-luminosity source, neither ground-based telescopes nor HSO/HIFI seem to be sensitive enough. The interferometer ALMA might provide in this case the unique possibility to constrain the HDO abundance.



\section{Conclusions} 
 
We reported theoretical predictions of the deuterated water
line emission from the envelopes of low mass protostars. 
In this study, we have focussed only on the envelope emission,
neglecting any HDO emission from the outflows, which is expected to be of minor importance based on theoretical and observational arguments.
We have shown that the simultaneous observations of appropriately 
selected transitions permit to approximatively constrain the HDO  
abundance in the outer cold and inner warm parts of the envelope. 
The uncertainty on the evaporation temperature mostly translates in 
an uncertainty on the inner abundance. The inner abundance can however
be quite accurately determined using the 225\,GHz line for which the emission has been 
shown to be independent from the evaporation temperature.

The most important result of the present study is that, for bright low-mass protostars (L$>$10\,L$_{\odot}$),
observations feasible from ground telescopes are enough to constrain the HDO 
abundance profile and evaporation temperature, and no HSO observations are required. 
In the case of low-luminosity protostars (L$<$10\,L$_{\odot}$),
both present ground based telescopes and HSO seem to lack sensitivity to observe HDO lines in a reasonable integration time. The HDO abundance profile and evaporation temperature will 
probably be directly constrained by future high resolution
observations with the large sub/millimeter interferometer ALMA.

\acknowledgements{We wish to thank the anonymous referee for very useful comments that helped
to improve this article.}


\bibliography{/Users/bparise/These/Manuscrit/biblio}
\bibliographystyle{aa}


\appendix

\section{The predicted fluxes for the testcase of IRAS16293$-$2422.}
\bigskip

\footnotesize
\begin{center}
  \tablefirsthead{\hline \hline \noalign{\smallskip}
Transition & Freq. & E$_{\rm up}$ & A$_{ul}$ & Line flux \\
           & {\tiny (GHz)}     & {\tiny (cm$^{-1}$)}  & {\tiny (s$^{-1}$)} & {\tiny (erg s$^{-1}$ cm$^{-2}$)} \\
\noalign{\smallskip}
\hline
\noalign{\smallskip}}
  \tabletail{\noalign{\smallskip} \hline}
  \tablehead{
\noalign{\smallskip}
\hline
\hline
\noalign{\smallskip}
\multicolumn{5}{l}{\small\it continued from previous page}\\
Transition & Freq. & E$_{\rm up}$ & A$_{ul}$ & Line flux \\
           & {\tiny (GHz)}    & {\tiny(cm$^{-1}$)}  & {\tiny (s$^{-1}$)} & {\tiny (erg s$^{-1}$ cm$^{-2}$)} \\
\noalign{\smallskip}
\hline
\noalign{\smallskip}
}
  \topcaption{Predicted fluxes for the IRAS16293$-$2422 testcase.}
  \label{fluxtest}
\scriptsize
\begin{supertabular}{ccccc}
$3_{3,0}-3_{3,1}$&0.824671&233.050&2.12$\times$10$^{-12}$&2.6$\times$10$^{-28}$\\
$5_{0,5}-4_{2,2}$&3.29676&221.946&4.34$\times$10$^{-13}$&3.2$\times$10$^{-27}$\\
$4_{3,1}-4_{3,2}$&5.70278&295.677&4.17$\times$10$^{-10}$&2.1$\times$10$^{-25}$\\
$2_{2,0}-2_{2,1}$&10.2782&109.269&3.63$\times$10$^{-9}$&1.3$\times$10$^{-22}$\\
$3_{2,1}-4_{1,4}$&20.4600&157.064&9.58$\times$10$^{-9}$&2.6$\times$10$^{-22}$\\
$5_{3,2}-5_{3,3}$&22.3077&374.409&1.64$\times$10$^{-8}$&1.7$\times$10$^{-23}$\\
$3_{2,1}-3_{2,2}$&50.2363&157.064&2.08$\times$10$^{-7}$&1.8$\times$10$^{-20}$\\
$4_{3,1}-5_{2,4}$&61.1859&295.676&2.66$\times$10$^{-7}$&1.3$\times$10$^{-21}$\\
$6_{0,6}-5_{2,3}$&69.5506&306.312&3.48$\times$10$^{-9}$&1.2$\times$10$^{-22}$\\
$1_{1,0}-1_{1,1}$&80.5783&32.4939&1.32$\times$10$^{-6}$&2.7$\times$10$^{-18}$\\
$5_{1,5}-4_{2,2}$&120.778&225.862&1.33$\times$10$^{-6}$&3.8$\times$10$^{-19}$\\
$6_{1,6}-5_{2,3}$&138.531&308.612&1.43$\times$10$^{-6}$&9.7$\times$10$^{-20}$\\
$4_{2,2}-4_{2,3}$&143.727&221.832&2.80$\times$10$^{-6}$&2.6$\times$10$^{-19}$\\
$3_{2,1}-4_{0,4}$&207.111&157.060&1.14$\times$10$^{-7}$&4.2$\times$10$^{-20}$\\
$3_{1,2}-2_{2,1}$&225.897&116.456&1.31$\times$10$^{-5}$&7.0$\times$10$^{-17}$\\
$2_{1,1}-2_{1,2}$&241.562&66.1781&1.18$\times$10$^{-5}$&6.5$\times$10$^{-17}$\\
$5_{2,3}-4_{3,2}$&255.050&303.989&1.78$\times$10$^{-5}$&1.0$\times$10$^{-18}$\\
$2_{2,0}-3_{1,3}$&266.161&109.262&1.75$\times$10$^{-5}$&1.0$\times$10$^{-17}$\\
$5_{2,3}-5_{2,4}$&310.533&303.987&1.78$\times$10$^{-5}$&1.1$\times$10$^{-18}$\\
$6_{2,5}-5_{3,2}$&313.751&384.867&3.75$\times$10$^{-5}$&1.1$\times$10$^{-18}$\\
$3_{3,1}-4_{2,2}$&335.396&233.016&2.61$\times$10$^{-5}$&1.2$\times$10$^{-18}$\\
$5_{3,2}-6_{1,5}$&356.836&374.402&3.53$\times$10$^{-7}$&5.8$\times$10$^{-21}$\\
$1_{0,1}-0_{0,0}$&464.925&15.4975&1.69$\times$10$^{-4}$&2.5$\times$10$^{-15}$\\
$3_{3,0}-4_{2,3}$&479.947&233.039&7.28$\times$10$^{-5}$&4.6$\times$10$^{-18}$\\
$3_{1,2}-3_{1,3}$&481.780&116.449&4.74$\times$10$^{-5}$&2.6$\times$10$^{-16}$\\
$2_{0,2}-1_{1,1}$&490.597&46.1612&5.25$\times$10$^{-4}$&6.4$\times$10$^{-15}$\\
$1_{1,0}-1_{0,1}$&509.292&32.4844&2.32$\times$10$^{-3}$&1.8$\times$10$^{-15}$\\
$2_{2,0}-3_{0,3}$&537.793&109.256&9.94$\times$10$^{-7}$&1.8$\times$10$^{-18}$\\
$2_{1,1}-2_{0,2}$&599.927&66.1706&3.45$\times$10$^{-3}$&2.9$\times$10$^{-15}$\\
$3_{1,2}-3_{0,3}$&753.411&116.444&5.90$\times$10$^{-3}$&4.2$\times$10$^{-15}$\\
$4_{1,3}-4_{1,4}$&797.487&182.965&1.31$\times$10$^{-4}$&3.2$\times$10$^{-16}$\\
$4_{1,3}-3_{2,2}$&827.263&182.964&1.28$\times$10$^{-3}$&4.2$\times$10$^{-15}$\\
$2_{1,2}-1_{1,1}$&848.962&58.1067&9.27$\times$10$^{-4}$&1.0$\times$10$^{-14}$\\
$1_{1,1}-0_{0,0}$&893.639&29.7880&8.35$\times$10$^{-3}$&1.1$\times$10$^{-14}$\\
$4_{3,1}-5_{1,4}$&912.605&295.656&2.74$\times$10$^{-6}$&2.1$\times$10$^{-19}$\\
$2_{0,2}-1_{0,1}$&919.311&46.1517&1.56$\times$10$^{-3}$&1.5$\times$10$^{-14}$\\
$4_{1,3}-4_{0,4}$&984.138&182.961&1.07$\times$10$^{-2}$&5.4$\times$10$^{-15}$\\
$3_{0,3}-2_{1,2}$&995.411&91.3064&7.04$\times$10$^{-3}$&2.3$\times$10$^{-14}$\\
$2_{1,1}-1_{1,0}$&1009.94&66.1608&1.56$\times$10$^{-3}$&1.5$\times$10$^{-14}$\\
$5_{2,3}-5_{1,4}$&1161.95&303.968&2.19$\times$10$^{-2}$&1.6$\times$10$^{-15}$\\
$4_{2,2}-4_{1,3}$&1164.77&221.809&2.00$\times$10$^{-2}$&3.0$\times$10$^{-15}$\\
$5_{1,4}-5_{1,5}$&1180.32&265.208&2.97$\times$10$^{-4}$&2.3$\times$10$^{-16}$\\
$3_{2,1}-3_{1,2}$&1217.26&157.036&1.89$\times$10$^{-2}$&4.4$\times$10$^{-15}$\\
$3_{1,3}-2_{1,2}$&1267.04&100.361&3.91$\times$10$^{-3}$&2.8$\times$10$^{-14}$\\
$2_{1,2}-1_{0,1}$&1277.68&58.0972&2.19$\times$10$^{-2}$&3.1$\times$10$^{-14}$\\
$2_{2,0}-2_{1,1}$&1291.64&109.239&1.59$\times$10$^{-2}$&6.3$\times$10$^{-15}$\\
$5_{1,4}-5_{0,5}$&1297.81&265.206&1.97$\times$10$^{-2}$&5.4$\times$10$^{-15}$\\
$3_{0,3}-2_{0,2}$&1353.78&91.2989&5.30$\times$10$^{-3}$&3.6$\times$10$^{-14}$\\
$3_{2,2}-2_{2,1}$&1392.92&155.357&3.25$\times$10$^{-3}$&7.3$\times$10$^{-15}$\\
$3_{2,1}-2_{2,0}$&1432.88&157.032&3.54$\times$10$^{-3}$&7.2$\times$10$^{-15}$\\
$5_{1,4}-4_{2,3}$&1444.83&265.202&1.02$\times$10$^{-2}$&8.6$\times$10$^{-15}$\\
$4_{0,4}-3_{1,3}$&1491.93&150.121&3.07$\times$10$^{-2}$&4.0$\times$10$^{-14}$\\
$3_{3,0}-4_{1,3}$&1500.99&233.016&3.67$\times$10$^{-6}$&7.9$\times$10$^{-19}$\\
$3_{1,2}-2_{1,1}$&1507.26&116.426&6.58$\times$10$^{-3}$&3.6$\times$10$^{-14}$\\
$2_{2,1}-2_{1,2}$&1522.93&108.890&2.06$\times$10$^{-2}$&1.1$\times$10$^{-14}$\\
$6_{1,5}-6_{1,6}$&1615.63&362.469&5.84$\times$10$^{-4}$&1.5$\times$10$^{-16}$\\
$3_{1,3}-2_{0,2}$&1625.41&100.353&4.49$\times$10$^{-2}$&5.4$\times$10$^{-14}$\\
$3_{2,2}-3_{1,3}$&1648.80&155.350&3.08$\times$10$^{-2}$&1.1$\times$10$^{-14}$\\
$4_{1,4}-3_{1,3}$&1678.58&156.343&9.92$\times$10$^{-3}$&4.1$\times$10$^{-14}$\\
$6_{1,5}-6_{0,6}$&1684.61&362.468&3.47$\times$10$^{-2}$&4.3$\times$10$^{-15}$\\
$4_{0,4}-3_{0,3}$&1763.56&150.115&1.20$\times$10$^{-2}$&5.0$\times$10$^{-14}$\\
$4_{2,3}-4_{1,4}$&1818.53&217.000&4.12$\times$10$^{-2}$&1.0$\times$10$^{-14}$\\
$4_{2,3}-3_{2,2}$&1848.31&216.999&1.06$\times$10$^{-2}$&1.1$\times$10$^{-14}$\\
$4_{3,2}-3_{3,1}$&1872.61&295.443&6.46$\times$10$^{-3}$&9.6$\times$10$^{-16}$\\
$4_{3,1}-3_{3,0}$&1877.49&295.634&6.51$\times$10$^{-3}$&9.7$\times$10$^{-16}$\\
$2_{2,1}-2_{0,2}$&1881.29&108.883&1.21$\times$10$^{-4}$&7.6$\times$10$^{-16}$\\
$3_{2,2}-3_{0,3}$&1920.43&155.344&3.04$\times$10$^{-4}$&1.1$\times$10$^{-15}$\\
$4_{2,2}-3_{2,1}$&1941.80&221.791&1.23$\times$10$^{-2}$&1.1$\times$10$^{-14}$\\
$5_{3,3}-6_{1,6}$&1950.15&373.620&1.11$\times$10$^{-5}$&9.4$\times$10$^{-19}$\\
$5_{3,3}-6_{1,6}$&1950.15&373.620&1.11$\times$10$^{-5}$&6.3$\times$10$^{-14}$\\
$5_{0,5}-4_{1,4}$&1965.55&221.900&8.28$\times$10$^{-2}$&4.2$\times$10$^{-14}$\\
$4_{1,3}-3_{1,2}$&1994.29&182.937&1.66$\times$10$^{-2}$&4.0$\times$10$^{-14}$\\
$4_{2,3}-4_{0,4}$&2005.18&216.995&5.46$\times$10$^{-4}$&8.0$\times$10$^{-16}$\\
$5_{3,3}-6_{0,6}$&2019.13&373.618&4.52$\times$10$^{-4}$&3.9$\times$10$^{-17}$\\
$5_{2,4}-5_{1,5}$&2031.74&293.589&5.43$\times$10$^{-2}$&7.9$\times$10$^{-15}$\\
$6_{1,5}-5_{2,4}$&2064.69&362.459&4.07$\times$10$^{-2}$&9.5$\times$10$^{-15}$\\
$5_{1,5}-4_{1,4}$&2083.04&225.816&1.98$\times$10$^{-2}$&3.4$\times$10$^{-14}$\\
$4_{3,2}-5_{1,5}$&2087.23&295.438&9.50$\times$10$^{-6}$&1.6$\times$10$^{-18}$\\
$5_{3,2}-5_{2,3}$&2110.99&374.360&8.12$\times$10$^{-2}$&3.0$\times$10$^{-15}$\\
$5_{2,4}-5_{0,5}$&2149.22&293.587&8.51$\times$10$^{-4}$&4.5$\times$10$^{-16}$\\
$5_{0,5}-4_{0,4}$&2152.20&221.896&2.23$\times$10$^{-2}$&4.0$\times$10$^{-14}$\\
$4_{3,2}-5_{0,5}$&2204.71&295.436&3.39$\times$10$^{-4}$&6.0$\times$10$^{-17}$\\
$4_{3,1}-4_{2,2}$&2213.71&295.626&7.73$\times$10$^{-2}$&5.3$\times$10$^{-15}$\\
$5_{1,5}-4_{0,4}$&2269.69&225.812&1.36$\times$10$^{-1}$&5.3$\times$10$^{-14}$\\
$3_{3,0}-3_{2,1}$&2278.02&232.998&5.79$\times$10$^{-2}$&7.5$\times$10$^{-15}$\\
$6_{2,5}-6_{1,6}$&2286.21&384.822&7.15$\times$10$^{-2}$&5.4$\times$10$^{-15}$\\
$2_{2,1}-1_{1,0}$&2291.31&108.873&1.26$\times$10$^{-1}$&3.9$\times$10$^{-14}$\\
$5_{2,4}-4_{2,3}$&2296.25&293.583&2.32$\times$10$^{-2}$&9.6$\times$10$^{-15}$\\
$3_{3,1}-4_{1,4}$&2297.65&232.970&5.62$\times$10$^{-6}$&1.8$\times$10$^{-18}$\\
$3_{3,1}-3_{2,2}$&2327.43&232.970&6.09$\times$10$^{-2}$&7.9$\times$10$^{-15}$\\
$5_{3,3}-4_{3,2}$&2343.74&373.612&1.89$\times$10$^{-2}$&1.7$\times$10$^{-15}$\\
$4_{3,2}-4_{2,3}$&2351.73&295.432&8.88$\times$10$^{-2}$&6.0$\times$10$^{-15}$\\
$6_{2,5}-6_{0,6}$&2355.19&384.820&1.23$\times$10$^{-3}$&2.4$\times$10$^{-16}$\\
$5_{3,2}-4_{3,1}$&2360.34&374.355&1.93$\times$10$^{-2}$&1.8$\times$10$^{-15}$\\
$2_{2,0}-1_{1,1}$&2382.17&109.214&1.29$\times$10$^{-1}$&4.4$\times$10$^{-14}$\\
$5_{3,3}-5_{2,4}$&2399.22&373.610&1.07$\times$10$^{-1}$&4.2$\times$10$^{-15}$\\
$6_{0,6}-5_{1,5}$&2411.83&306.258&1.70$\times$10$^{-1}$&3.3$\times$10$^{-14}$\\
$5_{2,3}-4_{2,2}$&2463.05&303.938&2.89$\times$10$^{-2}$&1.0$\times$10$^{-14}$\\
$5_{1,4}-4_{1,3}$&2465.87&265.179&3.27$\times$10$^{-2}$&2.7$\times$10$^{-14}$\\
$6_{1,6}-5_{1,5}$&2480.81&308.558&3.44$\times$10$^{-2}$&2.0$\times$10$^{-14}$\\
$3_{3,1}-4_{0,4}$&2484.30&232.966&1.56$\times$10$^{-4}$&5.1$\times$10$^{-17}$\\
$6_{0,6}-5_{0,5}$&2529.31&306.256&3.67$\times$10$^{-2}$&2.3$\times$10$^{-14}$\\
$6_{1,6}-5_{0,5}$&2598.29&308.556&2.18$\times$10$^{-1}$&3.7$\times$10$^{-14}$\\
$3_{2,2}-2_{1,1}$&2674.28&155.327&1.59$\times$10$^{-1}$&4.8$\times$10$^{-14}$\\
$6_{2,5}-5_{2,4}$&2735.28&384.812&4.22$\times$10$^{-2}$&6.9$\times$10$^{-15}$\\
$2_{2,0}-1_{0,1}$&2810.88&109.204&2.43$\times$10$^{-4}$&2.2$\times$10$^{-15}$\\
$6_{1,5}-5_{1,4}$&2916.11&362.440&5.53$\times$10$^{-2}$&1.5$\times$10$^{-14}$\\
$3_{2,1}-2_{1,2}$&2966.08&156.995&1.62$\times$10$^{-1}$&5.8$\times$10$^{-14}$\\
$4_{2,3}-3_{1,2}$&3015.33&216.972&2.05$\times$10$^{-1}$&4.6$\times$10$^{-14}$\\
$5_{3,3}-5_{1,4}$&3250.64&373.591&7.43$\times$10$^{-4}$&9.9$\times$10$^{-17}$\\
$5_{2,4}-4_{1,3}$&3317.29&293.559&2.63$\times$10$^{-1}$&3.6$\times$10$^{-14}$\\
$3_{2,1}-2_{0,2}$&3324.45&156.988&8.77$\times$10$^{-4}$&4.6$\times$10$^{-15}$\\
$4_{3,2}-4_{1,3}$&3372.77&295.409&4.06$\times$10$^{-4}$&1.0$\times$10$^{-16}$\\
$3_{3,1}-3_{1,2}$&3494.45&232.943&1.57$\times$10$^{-4}$&6.6$\times$10$^{-17}$\\
$6_{2,5}-5_{1,4}$&3586.70&384.793&3.37$\times$10$^{-1}$&2.4$\times$10$^{-14}$\\
$4_{2,2}-3_{1,3}$&3640.83&221.751&1.96$\times$10$^{-1}$&5.9$\times$10$^{-14}$\\
$3_{3,1}-2_{2,0}$&3710.07&232.938&6.26$\times$10$^{-1}$&3.6$\times$10$^{-14}$\\
$3_{3,0}-2_{2,1}$&3721.17&232.965&6.30$\times$10$^{-1}$&3.5$\times$10$^{-14}$\\
$4_{2,2}-3_{0,3}$&3912.47&221.746&1.99$\times$10$^{-3}$&4.4$\times$10$^{-15}$\\
$3_{3,0}-3_{1,3}$&3977.05&232.958&1.37$\times$10$^{-4}$&5.8$\times$10$^{-17}$\\
$4_{3,2}-3_{2,1}$&4149.80&295.391&6.85$\times$10$^{-1}$&3.4$\times$10$^{-14}$\\
$4_{3,1}-4_{1,4}$&4175.96&295.581&3.25$\times$10$^{-4}$&8.7$\times$10$^{-17}$\\
$4_{3,1}-3_{2,2}$&4205.74&295.580&7.02$\times$10$^{-1}$&3.3$\times$10$^{-14}$\\
$3_{3,0}-3_{0,3}$&4248.69&232.953&2.23$\times$10$^{-3}$&8.7$\times$10$^{-16}$\\
$4_{3,1}-4_{0,4}$&4362.61&295.576&8.17$\times$10$^{-3}$&2.0$\times$10$^{-15}$\\
$5_{2,3}-4_{1,4}$&4425.31&303.892&2.21$\times$10$^{-1}$&4.9$\times$10$^{-14}$\\
$5_{3,2}-5_{1,5}$&4453.27&374.306&5.25$\times$10$^{-4}$&8.5$\times$10$^{-17}$\\
$5_{3,3}-4_{2,2}$&4551.74&373.561&7.54$\times$10$^{-1}$&3.0$\times$10$^{-14}$\\
$5_{3,2}-5_{0,5}$&4570.75&374.304&1.73$\times$10$^{-2}$&2.5$\times$10$^{-15}$\\
$5_{2,3}-4_{0,4}$&4611.96&303.888&3.35$\times$10$^{-3}$&2.6$\times$10$^{-15}$\\
$5_{3,2}-4_{2,3}$&4717.77&374.300&7.96$\times$10$^{-1}$&2.9$\times$10$^{-14}$\\
$3_{3,0}-2_{1,1}$&5002.54&232.935&4.68$\times$10$^{-4}$&1.8$\times$10$^{-16}$\\
$3_{3,1}-2_{1,2}$&5243.27&232.902&4.16$\times$10$^{-4}$&1.5$\times$10$^{-16}$\\
$4_{3,1}-3_{1,2}$&5372.76&295.553&1.32$\times$10$^{-3}$&3.3$\times$10$^{-16}$\\
$3_{3,1}-2_{0,2}$&5601.64&232.894&3.24$\times$10$^{-3}$&7.3$\times$10$^{-16}$\\
$5_{3,2}-4_{1,3}$&5738.82&374.277&2.85$\times$10$^{-3}$&4.7$\times$10$^{-16}$\\
$4_{3,2}-3_{1,3}$&5848.84&295.351&1.02$\times$10$^{-3}$&1.8$\times$10$^{-16}$\\
$4_{3,2}-3_{0,3}$&6120.47&295.346&1.56$\times$10$^{-2}$&1.5$\times$10$^{-15}$\\
$5_{3,3}-4_{1,4}$&6514.00&373.515&1.83$\times$10$^{-3}$&1.8$\times$10$^{-16}$\\
$5_{3,3}-4_{0,4}$&6700.65&373.511&4.38$\times$10$^{-2}$&2.7$\times$10$^{-15}$\\
\end{supertabular}
\end{center}

\normalsize
\section{The low luminosity case : L1448mm}
\label{l14}
We present here the results for the case of the low-luminosity source L1448mm.
The density and temperature profiles where derived by \citet{Jorgensen02}.

\begin{figure} [!htbp]
\includegraphics[angle=0,width=1.\columnwidth]{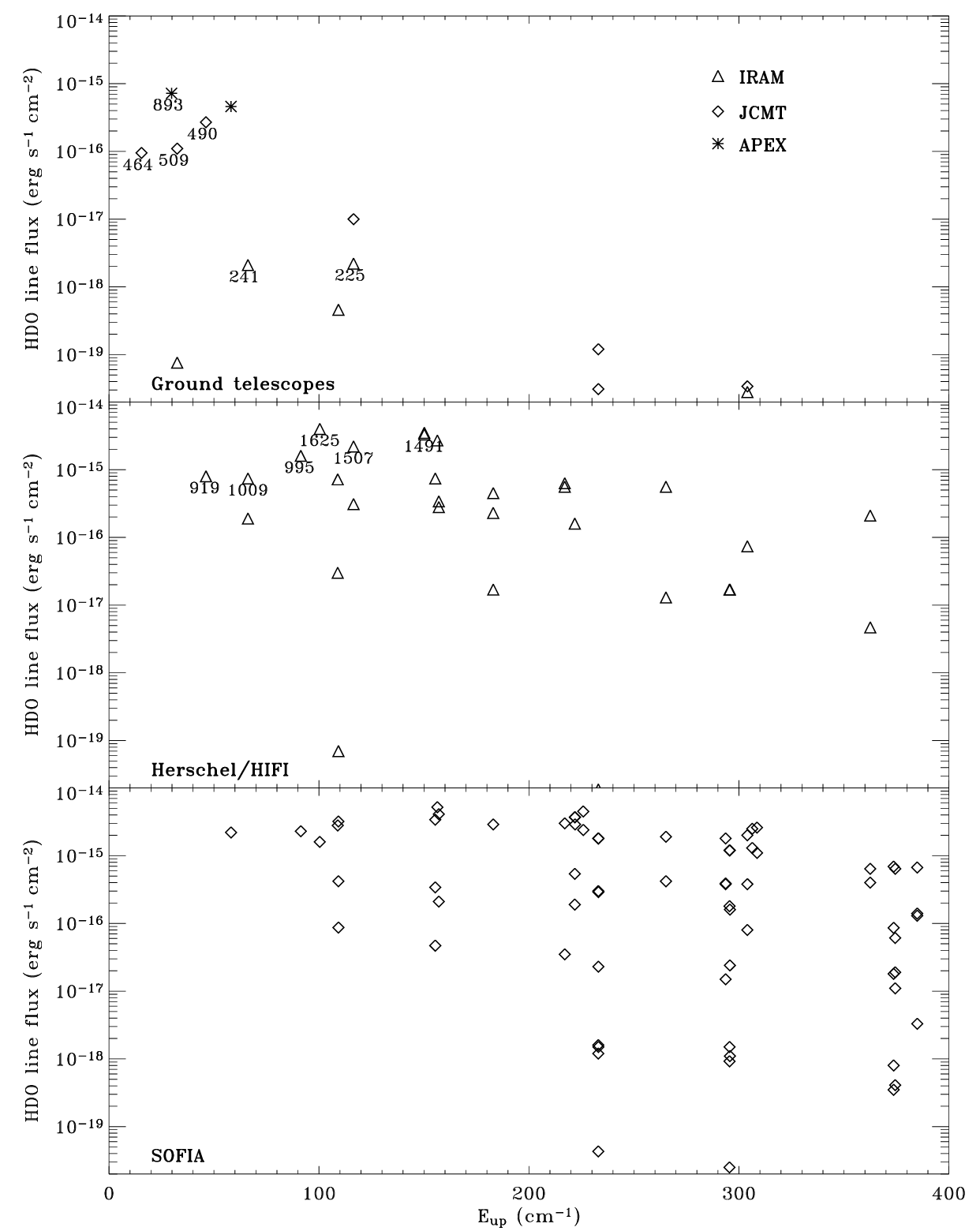} 
\caption{HDO line fluxes predicted for the case of the low-luminosity source L1448mm,
for an assumed HDO abundance of 10$^{-10}$ and 10$^{-7}$ in the cold and warm parts of the envelope respectively. The upper panel presents the transitions observable from the ground, the middle panel the transitions observable with HSO/HIFI and the lower panel the transitions observable with SOFIA. The frequencies of the transitions that are going to be studied in more detail are indicated in GHz.
} 
\label{lin14} 
\end{figure}

\begin{figure} [!htbp]
\includegraphics[angle=0,width=1.\columnwidth]{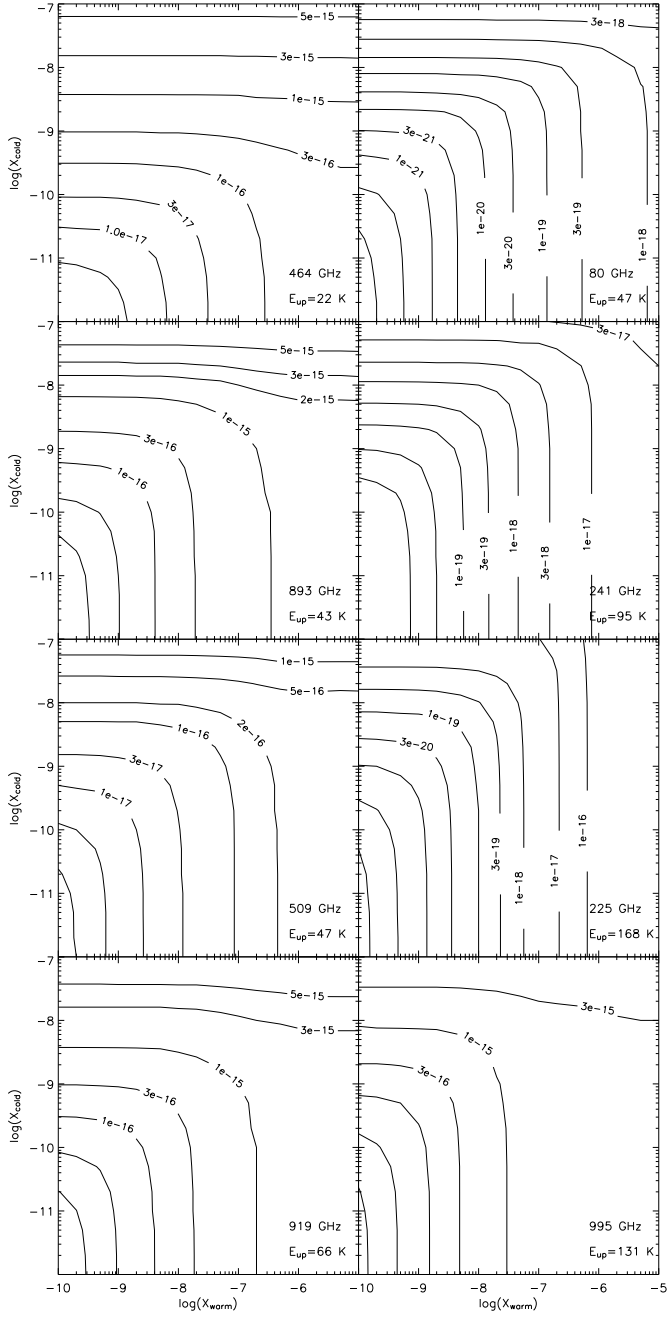} 
\caption{Line intensity  
as a function of the HDO abundance in the outer and inner regions 
of the envelope, $x_{\rm cold}$ and $x_{\rm warm}$ respectively, 
for the eight HDO transitions at 464, 893, 509, 80, 241, 225, 919 and 995 
GHz respectively towards the low-luminosity source L1448mm.
Line fluxes are in erg\,s$^{-1}$\,cm$^{-2}$. 
In these calculations the mantle evaporation temperature is 100\,K. 
} 
\label{coolvswarm_l14} 
\end{figure} 

\begin{figure} [!htbp]
\includegraphics[angle=0,width=1.\columnwidth]{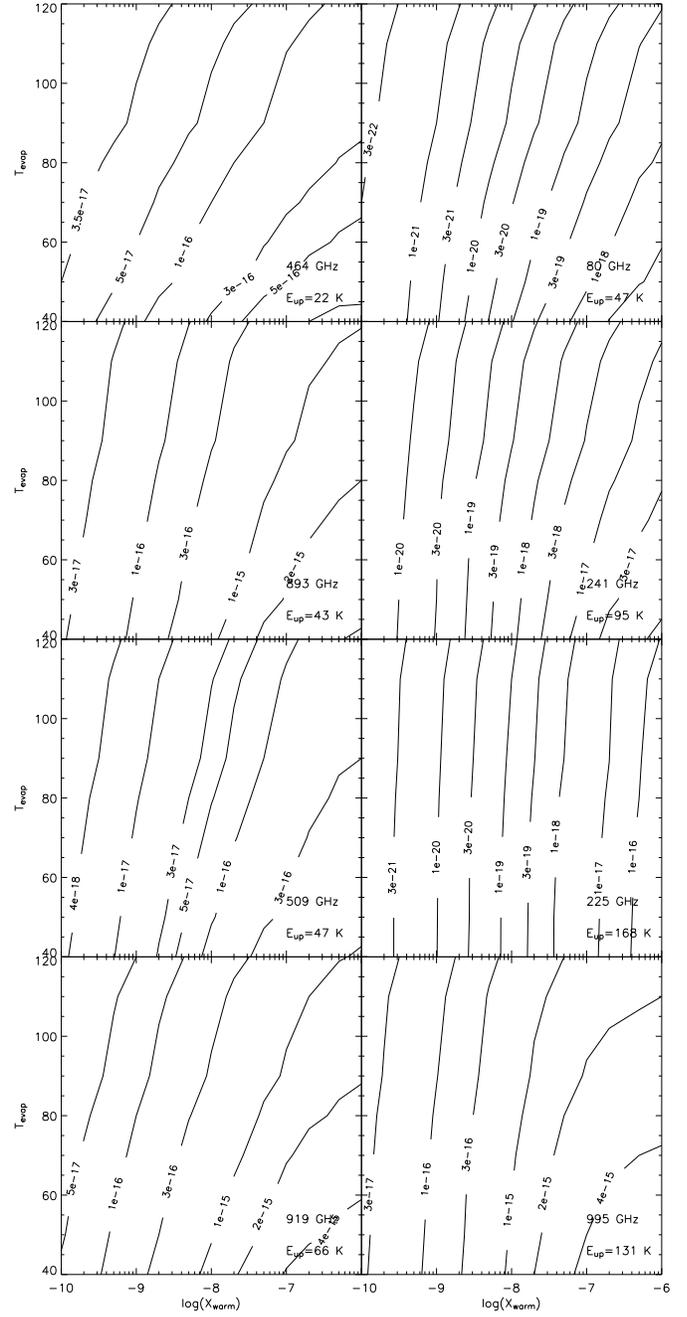} 
\caption{Line intensity  
as a function of the inner HDO abundance and the evaporation temperature,
for the eight HDO transitions at 464, 893, 509, 80, 241, 225, 919 and 995 
GHz respectively towards the low-luminosity source L1448mm.
Line fluxes are in erg\,s$^{-1}$\,cm$^{-2}$. 
In these calculations the outer HDO abundance is 10$^{-10}$. 
} 
\label{xinvstemp_l14} 
\end{figure}

\end{document}